\documentclass{jetpl}
\usepackage{color,amsmath,amsfonts,graphics,epsfig,amssymb}
\twocolumn

\lat

\title{Estimates of the cosmic gamma-ray flux at PeV to EeV energies from
the EAS-MSU experiment data }

\rtitle{Estimates of the cosmic gamma-ray flux
\ldots}

\sodtitle{Estimates of the cosmic gamma-ray flux at PeV to EeV energies from
the EAS-MSU experiment data }

\author{Yu.\,A.\,Fomin$^{1}$,
N.\,N.\,Kalmykov$^{1}$,
G.\,V.\,Kulikov$^{1}$,
V.\,P.\,Sulakov$^{1}$
and
S.\,V.\,Troitsky$^{2}$\thanks{e-mail: st@ms2.inr.ac.ru}}

\rauthor{Yu.\,A.\,Fomin et al.}

\sodauthor{Yu.\,A.\,Fomin,
N.\,N.\,Kalmykov,
G.\,V.\,Kulikov,
V.\,P.\,Sulakov
and
S.\,V.\,Troitsky}

\address{
$^{1}$~D.V.~Skobeltsyn Institute of Nuclear Physics, \\
M.V.~Lomonosov Moscow State University, Moscow 119991, Russia\\
$^{2}$~Institute for Nuclear Research of the Russian Academy of
Sciences,\\
60th October Anniversary prospect 7A, 117312 Moscow, Russia}

\dates{October 9, 2014}{*}

\abstract{
Archival EAS--MSU data are searched for anomalous muonless events which
may be caused by primary gamma rays with energies between $10^{15}$~eV and
$10^{18}$~eV. We
consider a refined sample of high-quality data and
confirm the previously reported detection of a non-zero
gamma-ray flux at $\sim 5\times 10^{16}$~eV
with a similar flux value but at somewhat lower statistical significance,
corresponding to a depletion of the sample. We
present upper limits on the flux below and
above these energies, including the first constraints in the range
$\left(10^{17} - 10^{18} \right)$~eV never studied by any other
experiment.
}

\PACS{98.70.Sa, 95.85.Pw, 98.70.Pw}

\begin{document}

\maketitle

 Searches for primary gamma rays in the extensive air shower (EAS) data
have continued since 1960s (see e.g.\ Ref.~\cite{KhristiansenBook} and, for
a recent review, Ref.~\cite{RisseReview}) but started to attract a special
attention after the recent announcement of the discovery of high-energy
astrophysical neutrinos by the IceCube collaboration \cite{IceCube1,
IceCube2, IceCube3}. Indeed, in conventional scenarios, the high-energy
neutrinos are produced in decays of charged pions, while accompanying
neutral pions decay into photons (see e.g.\ Refs.~\cite{Gupta-gamma,
Anch-accGamma, Kohta, KT} for more detailed discussions). Discovery of the
neutrinos thus gives a hope to find the accompanying photons.

One of the most elaborated methods to discriminate primary gamma rays is
to search for air showers with low muon content because secondary muons
are produced in hadronic interactions while the photon-induced EAS are
mostly electromagnetic. A recent study of EAS-MSU events with estimated
number of particles $N_{e}>2 \times 10^{7}$ has revealed \cite{3} an
excess
of muonless events compatible with non-zero primary gamma-ray flux at
energies $\gtrsim 50$~PeV. The present work
verifies the reported value of the flux with a refined sample of
high-quality data and
extends the $N_{e}$ range to
demonstrate a coherent picture of the photon-flux measurements and upper
limits in a wide energy band between $\sim 5$~PeV and $\sim 500$~PeV.

The EAS-MSU experiment~\cite{EAS-MSU} and the analysis method~\cite{3} are
described in detail in previous works. For the present study, events with
the reconstructed number of particles $N_{e}>10^{6}$ and zenith angles
$\theta<30^{\circ}$, detected in 1982--1990, are selected. Muons with
energies $>10$~GeV were recorded by the central muon detector of
36.4~m$^{2}$ area. For air showers with $N_{e}>10^{7}$, the triggering and
selection systems have been described in Ref.~\cite{3}. At lower $N_{e}$,
the central selection system was used, based on a subset of 7 scintillator
detectors; the central one, of 1~m$^{2}$ area, and 6 peripheric ones, each
of 0.5~m$^{2}$ area, located at $\sim 60$~m from the central one. The
trigger corresponds to a simultaneous (within the time gate of 500~ns)
firing of the central detector (with the threshold of 1 relativistic
particle) and of at least two of peripheric ones (whose thresholds were
set at the level of 1/3 of a relativistic particle). It is required that
the 3 detectors do not lay on a straight line so that the determination of
the EAS arrival direction is possible. The time resolution of the system
is $\sim 5$~ns, which determines the precision of $\le 3^{\circ}$ in the
arrival direction.
The precision in determination of the core position, important for the
present study, is $\sim 5\%$ of the distance from the shower axis to the
installation center, that is does not exceed $\sim 12$~m
even for the farthest events in the sample.
The number of particles in a shower is determined by an array of
Geiger-Mueller counters with the accuracy of $\sim 15\%$.

\begin{figure}
\includegraphics[width=0.9\columnwidth]{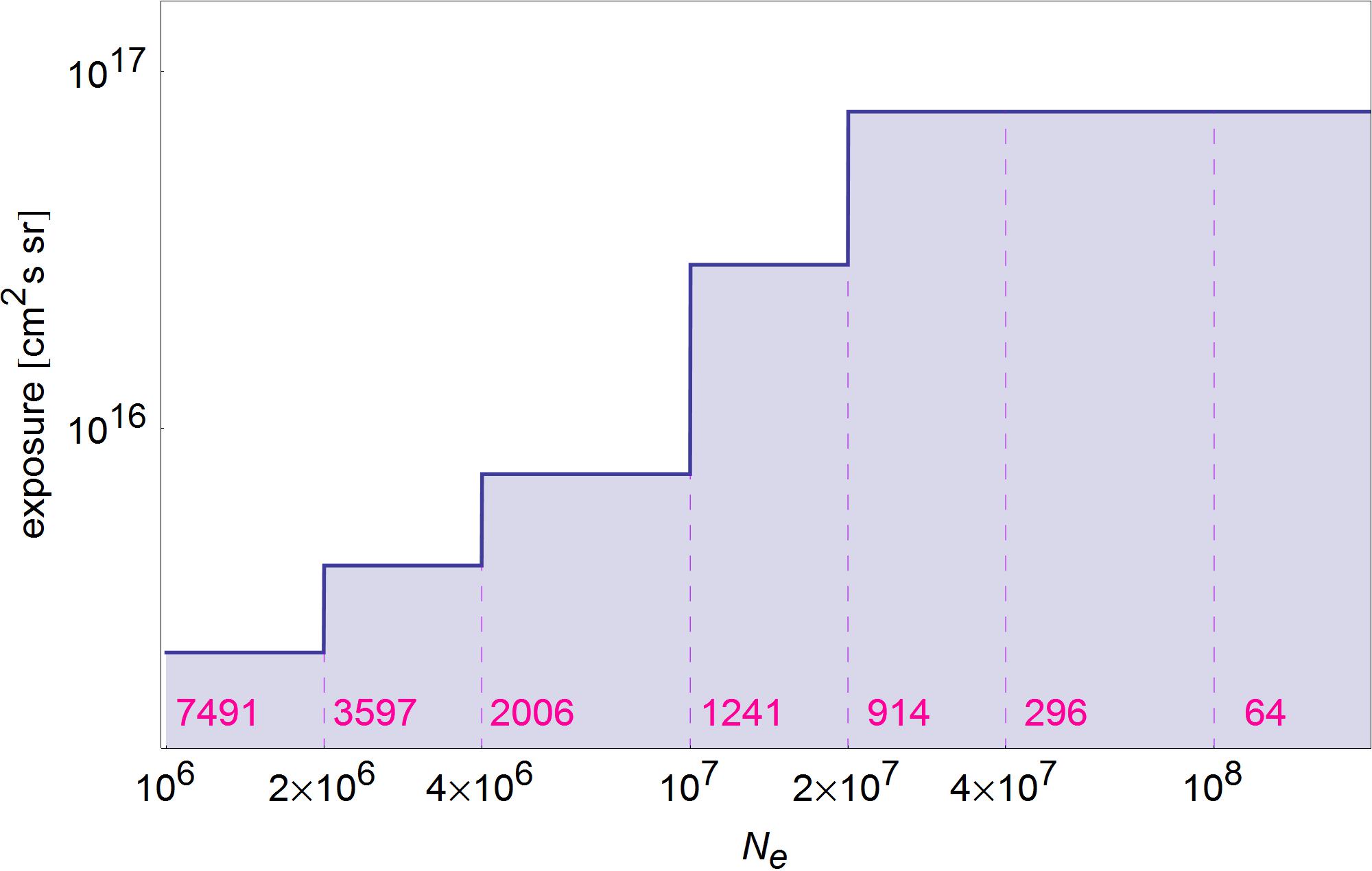}
\caption{\label{fig:exposure}
Figure~\ref{fig:exposure}.
The $N_{e}$ dependence of the experiment's exposure. Numbers indicate the
total amount of events in the sample within the corresponding $N_{e}$ bin.
}
\end{figure}

In each interval in $N_{e}$, the showers in the sample were selected by
their core position in such a way that the probability to register an
event is not less that 95\%. This means that the effective exposure
changes with $N_{e}$, though remains constant within every $N_{e}$ bin.
For the estimates of the integral flux, which require information from
several bins, we choose to select constant-exposure subsamples of the
available data to exclude the dependence of the result from the assumed
gamma-ray spectrum. At $N_{e}\ge 2\times 10^{7}$, the outer selection
system is used (see Ref.~\cite{3}) and the exposure does not change any
longer while the $\ge 95\%$ efficiency is kept.
To remove potential instrumental origin of the excess of photon-like
events, a careful check of the observation history was performed and the
days with any irregularities, such as a detector failure, were removed
from the sample. This reduces the exposure at large $N_{e}$ by $\sim1/3$
as compared to the previous study.
The exposure as a function
of $N_{e}$ is presented in Fig.~\ref{fig:exposure}.

An event is considered  a photon candidate if the 36.4~m$^{2}$ muon
detector located in the array's center did not record any signal.
These muonless showers, however, may
be rarely produced by primary hadrons. The method to estimate the expected
background of muonless events from hadronic showers was described in
detail in Ref.~\cite{3}. It includes simulations of artificial
proton-induced showers by means of the AIRES~2.6.0 \cite{AIRES} package
with the QGSJET-01 \cite{QGSJET} hadronic interaction model.

\begin{figure}
\includegraphics[width=0.9\columnwidth]{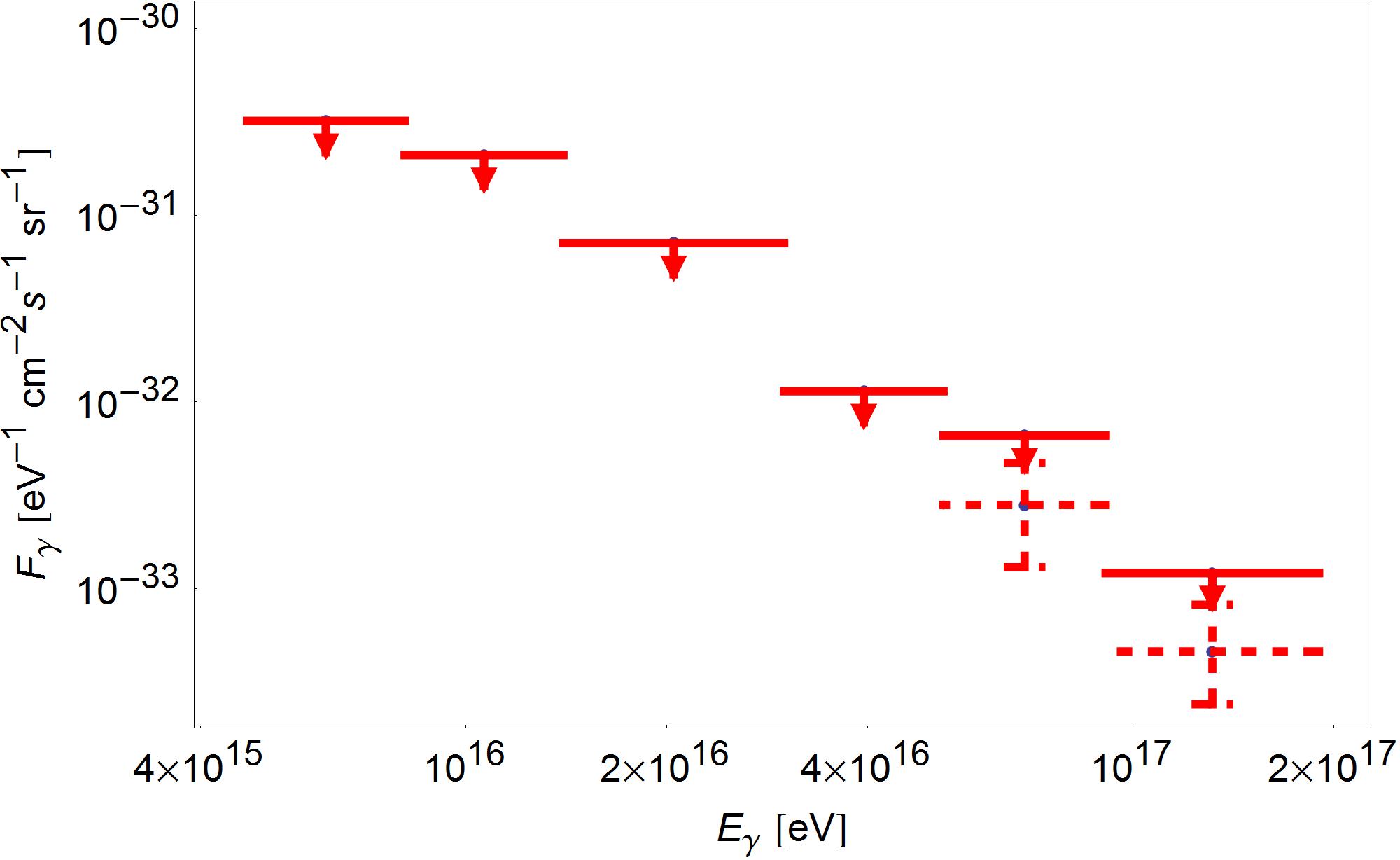}
\caption{\label{fig:diff-flux}
Figure~\ref{fig:diff-flux}.
The differential diffuse gamma-ray flux as estimated in this work, see
Table~\ref{tab:diff}. Horizontal bars indicate energy bins. For the two
highest-energy bins, both 95\% CL upper limits and 68\% CL error bars are
shown.}
\end{figure}

\begin{figure}
\includegraphics[width=0.9\columnwidth]{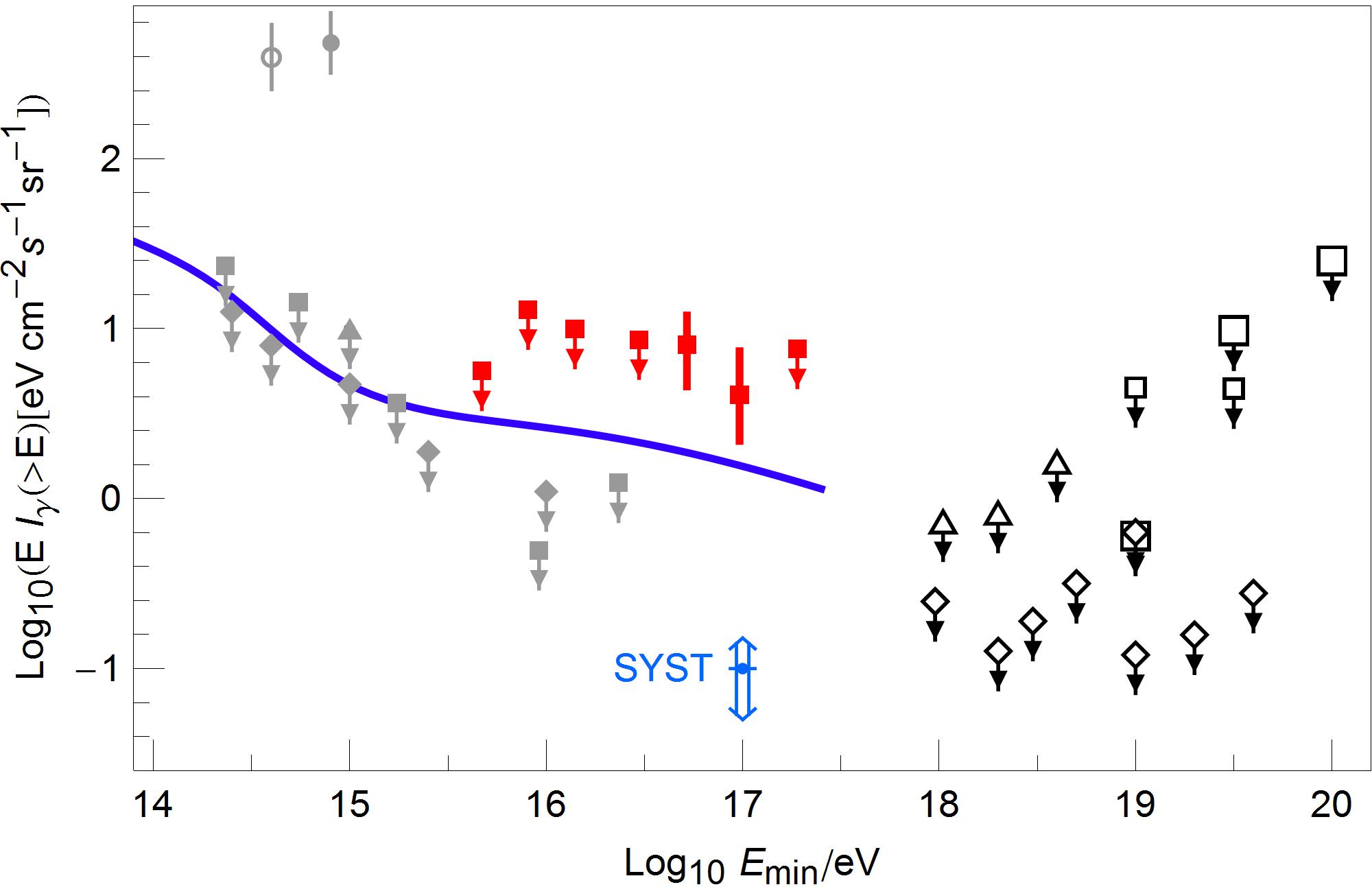}
\caption{\label{fig:int-flux}
Figure~\ref{fig:int-flux}.
Estimates of the integral gamma-ray flux from EAS-MSU (dark (red) squares
and error bars; this
work) and from other experiments: gray symbols -- open circle (Tien
Shan~\cite{Tien}, detection claim), circle (Lodz~\cite{Gawin}, detection
claim), triangles (EAS-TOP~\cite{EAS-TOP}), squares
(CASA-MIA~\cite{CASA-MIA}), diamonds (KASCADE~\cite{KASCADE, MMWG}); black
open symbols -- triangles (Yakutsk~\cite{Yak2}), diamonds (Pierre Auger
\cite{PAO1, PAO2}), small squares (AGASA~\cite{AGASA}), large squares
(Telescope Array~\cite{TA}). Systematic uncertainties (estimated for
EAS-MSU but relevant for other data as well, see text) are shown by a
double arrow. The curve represents a theoretical prediction of
Ref.~\cite{KT} for the model in which photons and neutrinos are produced
in cosmic-ray collisions with the hot gas surrounding our Galaxy, assuming
the best-fit IceCube observed neutrino spectrum.}
\end{figure}

The results of the bin-by-bin study are presented in Table~\ref{tab:diff}
\begin{table}
\begin{center}
\begin{tabular}{ccccc}
\hline
$N_{e}$, & $E_{\gamma} $, & $N_{\rm obs}$ & $N_{\rm exp}$
&  $10^{32}F_{\gamma}$,\\
$10^{6}$ & PeV            &               &
&
%$10^{-32}$~
(eV\,cm$^{2}$\,s\,sr)$^{-1}$\\
\hline
1 -- 2 & 4.7 -- 8.1 & 17 & 34.0 & $<$32\\
2 -- 4 & 8.1 -- 14 & 7 & 8.7 & $<$21\\
4 -- 10 & 14 -- 30 & 3 & 3.8 & $<$3.9\\
10 -- 20 & 30 -- 52 & 5 & 4.0 & $<$1.14\\
20 -- 40 & 52 -- 91 & 25 & 16.5 & $<$0.660 \\
40 -- 100 & 91 -- 190 & 4 & 0.5 & $<$0.121\\
          &               &   &     & $0.046^{+0.036}_{-0.022}$\\
\hline
\end{tabular}
\end{center}
\caption{\label{tab:diff}
Table~\ref{tab:diff}.
Estimates of the differential diffuse gamma-ray flux. $E_{\gamma}$ is the
mean energy of a primary photon which produces an EAS with the
corresponding $N_{e}$;   $N_{\rm obs}$ is the number of observed muonless
events in the bin; $N_{\rm exp}$ is the expected number of background
muonless events from usual cosmic rays; $F_{\gamma}$ is the estimate of
the differential gamma-ray flux in the bin. It does not depend on the
assumed photon spectrum because the exposure is constant within a bin.
}
\end{table}
and Fig.~\ref{fig:diff-flux}
while the estimate of the integral gamma-ray flux is given in
Table~\ref{tab:int}
\begin{table}
\begin{center}
\begin{tabular}{ccccc}
\hline
$N_{e}^{\rm min}$, & $E_{\gamma}^{\rm min}$, &
$N_{\rm obs}$ & $N_{\rm exp}$ &
$10^{16}I_{\gamma}$, \\
$10^{6}$           & PeV &&                    &
(cm$^{2}$\,s\,sr)$^{-1}$\\
\hline
1 & 4.7 & 19 & 34.0 & $<12$ \\
2 & 8.1 & 8 & 8.0 & $<$16 \\
4 & 14 & 4 & 4.5 &$<7.1$\\
10 & 30 & 6 & 4.3 &  $<2.9$\\
20 & 52 & 29 & 17.0 & $<3.1$\\
   &      & & & $1.55^{+0.75}_{-0.67}$\\
40 & 91 & 4 & 0.5 & $<$1.20\\
   &      & & & $0.45^{+0.36}_{-0.21}$\\
100 & 190 &0 & $<0.001$ & $<$0.40\\
\hline
\end{tabular}
\end{center}
\caption{\label{tab:int}
Table~\ref{tab:int}.
Estimates of the integral diffuse gamma-ray flux $I_{\gamma}$ at photon
energies above $E_{\gamma}^{\rm min}$, corresponding to $N_{e}>N_{e}^{\rm
min}$.
}
\end{table}
and compared to results of other experiments in Fig.~\ref{fig:int-flux}.
The latter plot presents also a comparison with an example theoretical
curve from Ref.~\cite{KT} normalized to the IceCube neutrino flux.
All
upper limits reported in the tables and plots are 95\% confidence level
(CL). Wherever a measurement with error bars is presented, the error bars
are 68\% CL, statistical only.

Several comments to these results are in order.

1. The excess of muonless events at $N_{e}\gtrsim 10^{7}$ suggests
\cite{3, Khorkhe} a nonzero gamma-ray flux at several dozen PeV.
In this work, we used a refined sample of high-quality data and obtained a
value of the flux in excellent agreement with the result reported in
\cite{3}. The statistical significance of this result is reduced in
accordance with the reduction in the used exposure by $\sim 1/3$.
Note that
the precise photon energy to which the flux corresponds is
slightly different from the previous work due to  more accurate estimates
of $N_{e}$ for gamma-ray showers (the
change is within the systematic uncertainties). The excess is not seen at
lower energies thus giving additional support to its physical, and not
instrumental, origin.
At energies above $\sim 10$~PeV, where the exposure is still
low, the excess starts to build up but its significance is low. Therefore,
the analysis carried out in several energy bins results in a coherent
picture supporting the first ever observation of cosmic gamma rays above
100~TeV with a relatively hard spectrum.

2. As it was discussed in Ref.~\cite{3}, the dominant source of systematic
errors for this kind of a study is related to the background estimation:
various models of hadronic interactions predict different muon content of
air showers; additional uncertainty is related to the chemical composition
of the bulk of primary cosmic rays. The resulting systematic error of the
derived gamma-ray flux is $\pm 50\%$ for the EAS-MSU data~\cite{3}, and a
similar scale of uncertainties is expected for other experiments whose
results are presented in Fig.~\ref{fig:int-flux}.
At lower energies of order a few PeV, the observed number of
muonless events is significantly lower than predicted in our
simulations, see Table~\ref{tab:diff}. The most probable reason for that
is the assumption of pure proton composition in the estimation of the
background of muonless events. With the realistic composition, the number
of muon-rich showers increases, thus reducing the background. We
however choose to keep the conservative high background for the detection
claim in higher energy bins and postpone a detailed estimate of the effect
of the realistic composition to a future work. The undercount of muonless
events at PeV energies is unlikely to be related to the excess of muons in
air showers with respect to simulations, as observed by the
Yakutsk~\cite{Yakutsk-mu} and Pierre Auger~\cite{PAOmuons, 1408.1421}
experiments at much higher primary energies but for lower-energy muons.

3. The constraints presented here for energies $\sim (10^{17} -
10^{18})$~eV cover the region unexplored by other experiments and present
therefore the first ever published limits on the diffuse gamma-ray flux at
these energies. These limits may be improved, or non-zero photon flux
observed, by future experiments like Tunka-HiSCORE~\cite{HiSCORE} or
low-energy extensions of the Telescope Array (TALE~\cite{TALE}) and Pierre
Auger (AMIGA~\cite{AMIGA}) experiments, as well as in reanalysis of
archival data of muon detectors in Yakutsk~\cite{Yakutsk-mu1}. Additional
tests of the origin of the (50--100)~PeV excess of muonless events in the
EAS-MSU data will be provided by a dedicated study of other EAS
observables which is currently under way.

We are indebted to Grigory Rubtsov for interesting and useful discussions
and to Oleg Kalashev for sharing the results of Ref.~\cite{KT} prior to
publication. The experimental work of the EAS-MSU group and computer
simulations were supported in part by the grants of the Government of the
Russian Federation (agreement 14.B25.31.0010) and of the Russian
Foundation for Basic Research (projects 14-02-00372 and 13-02-00214).
Development of the methods to search for primary photons and application
of these methods to the EAS-MSU data were
supported by the Russian Science Foundation, grant 14-12-01340.

\end{document}